\documentclass[11pt]{article}
\usepackage{graphicx}
\graphicspath{{C:/Users/xinwei/Documents/6EPMapping/gunterPaper/}}

\columnwidth\textwidth

\def\fii{\varphi}

\def\ro{\varrho}
\def\si{\sigma}

\def\d{\partial}
\def\=d{\,{\buildrel\rm def\over =}\,}

\def\sqr#1#2{{\vcenter{\vbox{\hrule height.#2pt\hbox{\vrule width.
#2pt height#1pt \kern#1pt \vrule width.#2pt}\hrule height.#2pt}}}}
 
\def\eps{\varepsilon}

\def\B{\Bigl}

\begin{document}

\title{Approach to steady state in the heat equation and in the hyperbolic heat transfer equation}

\author{G\"unter Scharf
\footnote{e-mail: scharf@physik.uzh.ch}
\\ Physics Institute, University of Z\"urich, 
\\ Winterthurerstr. 190 , CH-8057 Z\"urich, Switzerland}

\date{ }

\maketitle\vskip 3cm

\vskip 3cm

\begin{abstract}

We investigate the spherically symmetric 1D ablation problem. We show that the parabolic heat equation fails to describe the approach to steady state in infinite space. The hyperbolic equation shows an approach to steady state with a time constant given by the thermal relaxation time. However the infinite geometry is rather unphysical and gives rise to a so-called zero mode. Therefore we also consider the finite problem with a large boundary at constant temperature. Then both equations show approach to steady state, but only the hyperbolic equation seems to be physically correct for small times.
\end{abstract}

\newpage
\section{Introduction}

It is well known that the usual heat equation
$$\ro c{\d T\over\d t}=\kappa\triangle T+Q\eqno(1.1)$$
has the defect that heat can propagate instantaneously through space. In (1.1) $T$ is temperature, $t$ is time, $\triangle$ is the Laplace operator and $Q$ describes the heat generation. The constant $\kappa$ is the thermal conductivity, $\ro$ the mass density and $c$ the specific heat. Still the equation is widely used to simulate heat transfer problems ([1] and references given there). The reason for this causality defect is that the heat equation is first order in time. A second defect related to this is the restriction of the initial value problem. In the Cauchy problem for (1.1) only the temperature $T$ at time $t=0$ can be specified. But in experiments also the temporal derivative $\d T/\d t$ at $t=0$ must be adjusted to the experimental situation. This is possible in the hyperbolic heat transfer equation which is second order in time.

It is our aim to compare the two equations in an analytically solvable case of some practical interest. In the next section the 1D ablation problem in infinite space is solved for the heat equation and in sect.3 for the hyperbolic heat transfer equation. In sect.4 some special functions are discussed which appear in the solutions. In sect.5 we study the problem on a finite spherical volume. This is important because it turns out that the infinite
problem cannot be viewed as the limit of the finite problem. So for real applications only this finite problem is relevant.

\section{The 1D ablation problem according to the bio-heat equation}

Let us consider a spherical electrode of radius $r_0$ in an infinite medium with electrical conductivity $\si$. Assuming the second dispersive electrode at infinity with potential $V=0$ the potential in the medium is given by the simple solution of Laplace's equation
$$V(r)={V_0r_0\over r}\eqno(2.1)$$
where $V_0$ is the applied potential on the ablation electrode. The corresponding electric field strength is
$$E_r=-{\d V\over\d r}={V_0r_o\over r^2}\eqno(2.2)$$
and the heat generation
$$Q=\si{(V_0r_0)^2\over r^4}.\eqno(2.3)$$
With this constant heating we want to calculate the transient temperature $T(t,r)$ as solution of equation (1.1) which becomes
$${\d T\over\d t}={a\over r^2}{\d\over\d r}\B(r^2{\d T\over\d r}\B)+{\beta\over r^4},\quad a={\kappa\over\ro c},$$
where $\beta$ is the constant appearing in (2.3) divided by $\ro c$ and $a$ the thermal diffusivity.

We write the solution as
$$T(t,r)=T_1(r)+T_2(t,r),\eqno(2.4)$$
where $T_1(r)$ is the steady state solution satisfying
$${d\over dr}\B(r^2{dT_1\over dr}\B)=-{\beta\over ar^2}.\eqno(2.5)$$
Assuming a temperature $T_\infty$ at infinity we get
$$T_1(r)=T_\infty+{C_1\over r}-{b\over 2r^2}.\quad b={\beta\over a}\eqno(2.6)$$
The integration constant $C_1$ is fixed by the assumption of no heat flux at the electrode
$${dT_1\over dr}\vert_{r_0}=0,\eqno(2.7)$$
which is reasonable for a small electrode. This gives the following steady state solution
$$T_1(r)=T_\infty+{b\over r_0 r}-{b\over 2r^2}.\eqno(2.8)$$
The maximal temperature is found at the electrode, of course
$$T_1(r_0)=T_\infty+{b\over 2r_0^2}.\eqno(2.9)$$

The remaining homogeneous equation for $T_2$ is solved by separation of variables
$$T_2(t,r)=T_3(t)T_4(r).\eqno(2.10)$$
Then we have
$${1\over a}{\dot T_3\over T_3}={T_4''\over T_4}+{2\over r}{T_4'\over T_4}={\rm const.}=-\,k^2$$
where the dot means time derivative and the prime radial derivative. This yields
$$T_3(t)=C_3e^{-\,k^2at}\eqno(2.11)$$
and
$$r^2T_4''+2rT_4'+\,k^2r^2T_4=0.\eqno(2.12)$$
This last equation is a special case of Bessel's equation with the solution
$$T_4(r)={1\over r}\B(C_4\sin\,k r+C_5\cos\,k r\B).\eqno(2.13)$$
Now the general solution of our problem is given by
$$T(t,r)=T_\infty+{b\over r_0 r}-{b\over 2r^2}+{1\over r}\int\limits_0^\infty d\,k\,e^{-\,k^2at}\B(f_1(\,k)\sin\,k r+$$
$$+f_2(\,k)\cos\,k r\B).\eqno(2.14)$$

The unknown functions $f_1$ and $f_2$ in (2.14) are determined by the initial condition
$$T(0,r)=T_\infty,\eqno(2.15)$$
$${b\over 2r^2}-{b\over r_0r}={1\over r}\int\limits_0^\infty[f_1(\,k)\sin\,k r+f_2(\,k)\cos\,k r]d\,k.\eqno(2.16)$$
Here the l.h.side is only defined for $r>r_0$. To use the theorems on the {\it real} Fourier transform we continue the functions to $0<r<r_0$. The even function $b/2r^2$ is continued by the constant $b/2r_0^2$. The inverse Fourier transform then yields
$$f_1(\,k)={b\over\pi}\int\limits_{r_0}^\infty{\sin\,k r\over r}dr+{b\over\pi r_0^2}\int\limits_0^{r_0}r\sin\,k r\,dr=$$
$$={b\over 2}-{b\over\pi}{\rm Si}(\,k r_0)+{b\over\pi\,k^2r_0^2}(\sin\,k r_0-\,k r_0\cos\,k r_0)=$$
$$={b\over 2}+O(\,k r_0).\eqno(2.17)$$
Here {\rm Si} is the sine integral. For small electrode radius $r_0$ this gives the following contribution to (2.14)
$$\int\limits_0^\infty d\,k e^{-\,k^2at}f_1(\,k)\sin\,k r={b\over 4\sqrt{at}}\int\limits_0^\infty dy\,e^{-y^2/4}\sin{ry\over 2\sqrt{at}}=$$
$$={b\over 2\sqrt{at}}D_+\B({r\over 2\sqrt{at}}\B).\eqno(2.18)$$
Here $D_+(z)$ is the Dawson function (see Wikipedia and references given there).

The function $f_2(\,k)$ in (2.16) must clearly degenerate to a delta-distribution
$$f_2(\,k)=-{2b\over r_0}\delta(\,k).\eqno(2.19)$$
This contribution then cancels the $1/r$-term in (2.14) which is necessary for $t=0$. Then the final result is
$$T(t,r)=T_\infty-{b\over 2r^2}+{b\over 2\sqrt{at}}D_+\B({r\over 2\sqrt{at}}\B)+O(r_0).\eqno(2.20)$$
The Dawson function has the following asymptotic expansion
$$D_+(x)={2\over 2x}+{1\over 4x^3}+{3\over 8x^5}+\ldots$$
for $x\gg 1$. Using this in (2.20), the first term cancels the negative contribution so that the result for small times seems to be correct. It shows the expected rise of the temperature above $T_\infty$. However the Dawson function has a maximum at $x=0.924...$, $D(x)=0.541...$ and for smaller $x$ it decreases to 0. As a consequence for large $t$ the temperature (2.20) falls below $T_\infty$ which is completely wrong ! The reason for this disaster is simple: The heat equation is first order in time. Therefore only one condition, namely the initial condition at $t=0$ is at our disposal. The hyperbolic heat equation in the next section is second order in time, then we have two free constants of integration, so that we can get the right behavior of the solution at $t=0$ {\it and} $t=\infty$.

\section{Approach to steady state in the hyperbolic heat equation}

According to Cattaneo [2] and Vernotte [3] a better description of heat transfer is obtained by assuming a time delay $\tau$ between heat flux $q$ and temperature gradient
$$q(t+\tau,x)=-\kappa \nabla T(t.x).\eqno(3.1)$$
Expanding up to first order in $\tau$ we have
$$q+\tau{\d q\over\d t}=-\kappa\nabla T$$
and
$$\nabla q=-\tau{\d\over \d t}\nabla q-\nabla \kappa\nabla t$$
which is substituted into the energy conservation equation
$$-\nabla q(t,x)+Q(t,x)=\ro c{\d T(t,x)\over\d t}.\eqno(3.2)$$
For constant thermal conductivity $k$ this gives the hyperbolic heat equation
$${\d^2 T\over\d t^2}+{1\over\tau}{\d T\over\d t}-{\kappa\over\tau\ro c}\triangle T={1\over\ro c}{\d Q\over\d t}+{Q\over\tau\ro c}.\eqno(3.3)$$

For our 1D ablation problem this equation assumes the following form
$${\kappa\over\tau\ro c}{1\over r^2}{\d\over\d r}\B(r^2{\d\over\d r}T\B)+{d\over r^4}={1\over\tau}{\d T\over\d t}+{\d^2T\over\d t^2}.\eqno(3.4)$$
With another boundary condition this problem has been studied in [4]. These authors use the method of Laplace transform which becomes very complicated. By the method of the previous section we get the solution in much simpler form. We do not include a switching of the heat generation $Q$ by means of a Heaviside step function as in [4]. This would give an additional singular term $u_1(r)\delta(t)$. Then we have a so-called generalized Cauchy problem in the sense of distributions, which has been treated by Vladimirov [7]. The $\delta$-term then fixes the initial condition at $t=0$ as
$${\d T(t,r)\over\d t}\B\vert_{t=0}=u_1(r).$$
But in [4] the simple initial condition
$${\d T(t,r)\over\d t}\B\vert_{t=0}=0\eqno(3.5)$$
was used, this is a certain inconsistency. We consider the classical Cauchy problem where we have two initial conditions at $t=0$ for free.

As before we write the solution in the form (2.4) where the steady state solution $T_1(r)$ satisfies the equation
$${d\over dr}\B(r^2{dT_1\over dr}\B)=-{\beta\over ar^2}.\eqno(3.6)$$
We have the same steady state solution (2.8)
$$T_1(r)=T_\infty+{b\over r_0 r}-{b\over 2r^2}.\eqno(3.7)$$
But the homogeneous equation now reads
$${1\over r^2}\B(2r{\d T_2\over\d r}+r^2{\d ^2T_2\over\d r^2}\B)={1\over a}{\d T_2\over\d t}+\eps{\d ^2T_2\over\d t^2}\eqno(3.8)$$ 
with
$$a={\ro c\over\kappa},\quad \eps=a\tau.\eqno(3.9)$$
For $\eps=0$ we are back at the parabolic heat equation.

Again equation (3.8) is solved by separating the variables
$$T_2(t,r)=T_3(t)T_4(r)\eqno(3.10)$$
which yields
$${1\over a}{\dot T_3\over T_3}+\eps{\ddot T_3\over T_3}={T_4''\over T_4}+{2\over r}{T_4'\over T_4}=-\,k^2.\eqno(3.11)$$
The equation for $T_4$ is the same as before (2.12), but for $T_3$ we now have the second order equation
$$\eps\ddot T_3+{1\over a}\dot T_3=-\,k^2T_3.\eqno(3.12)$$
It has two exponential fundamental solutions
$$T_\pm (t)=C_\pm e^{\omega_\pm t}\eqno(3.13)$$
where $\omega_\pm$ are the two solutions of the quadratic equation
$$\eps\omega^2+{1\over a}\omega+\,k^2=0.$$
We have two negative roots
$$\omega_\pm=-{1\over 2a\eps}\pm\sqrt{{1\over 4a^2\eps^2}-{\,k^2\over\eps}}=-{1\over 2\tau}\pm\sqrt{{1\over 4\tau^2}-{a\,k^2\over\tau}},\eqno(3.14)$$
Then the general solution with the same boundary condition (2.7) as in the last section is given by
$$T(t,r)=T_\infty+{b\over r_0r}-{b\over 2r^2}+{1\over r}\int\limits_0^\infty d\,k e^{\omega_+t}\B[f_1(\,k)\sin \,k r+$$
$$+f_2(\,k)\cos\,k r\B]+{1\over r}\int\limits_0^\infty d\,k e^{\omega_-t}\B[g_1(\,k)\sin \,k r
+g_2(\,k)\cos\,k r\B].\eqno(3.15)$$

To satisfy the initial condition $T=T_\infty$ we must again compensate the second term $\beta/r_0r$ by some contribution from the integrals for $\,k =0$. For $\,k=0$ we have
$$\omega_+=0,\quad \omega_-=-{1\over\tau}.\eqno(3.16)$$
In the first case with $f_2(k)\sim\delta(\,k)$ we are in the same situation as in the last section and get no approach to steady state. So we take $f_2=0$. But now we can choose
$$g_2(\,k)=-{b\over r_0}\delta(\,k)\eqno(3.17)$$
and have the desired compensation for $t=0$. But for $t\to\infty$ this term goes to 0 because $\omega_-$ (3.16) gives an exponential fall off $\sim\exp -t/\tau$. That means we obtain the correct steady state as far as the $1/r$ term is concerned.

Regarding the $1/r^2$ term we must determine $f_1$ and $g_1$ such that
$${b\over r}=\int\limits_0^\infty d\,k\,[f_1(\,k)+g_1(\,k)]\sin\,k r.\eqno(3.18)$$
As before (2.17) this gives
$$f_1(\,k)+g_1(\,k)={b\over 2}+O(\,k r_0).\eqno(3.19)$$
To determine $f_1$ and $g_1$ separately we need a second initial condition. Preliminary experiments show that the above condition (3.5) is physically correct, so we assume it. The condition (3.5) implies
$$\omega_+f_1+\omega_- g_1=0$$
so that
$$f_1(\,k)={b\over 2}{\omega_-\over\omega_--\omega_+},\quad
g_1(\,k)={b\over 2}{\omega_+\over\omega_+-\omega_-}.\eqno(3.20)$$
Inserting the roots (3.14) we obtain the following final result
$$T(t,r)=T_\infty+{b\over r_0r}\B(1-e^{-t/\tau}\B)-{b\over 2r^2}+$$
$$+{b\over 4r}\int\limits_0^\infty d\,k\B[e^{\omega_+t}\B(1+(1-4a\tau\,k^2)^{-1/2}\B)+e^{\omega_-t}\B(1-(1-4a\tau\,k^2)^{-1/2}\B)\B]\sin\,k r.\eqno(3.21)$$
However, we note that this total solution does not satisfy the initial condition (3.5) of vanishing temporal derivative. We shall return to this point in sect.5.

The integral in (3.21) must be split at
$$\,k={1\over 2\sqrt{a\tau}}=\,k_0\eqno(3.22)$$
because the roots (3.14) become complex for $k>k_0$. For small $\tau$ only the integral
$$I_1=\int\limits_0^{\,k_0}d\,k\B(1+(1-4a\tau\,k^2)^{-1/2}\B)e^{\omega_+t}\sin\,k r\eqno(3.23)$$
is important. Extending the upper limit to infinity and expanding the square root we get
$$I_1=\int\limits_0^\infty d\,k(2+2a\tau\,k^2)e^{-at\,k^2}\sin\,k r=$$
$$=2(1-\tau\d_t)\int\limits_0^\infty d\,k\,e^{-at\,k^2}\sin\,k r.\eqno(3.24)$$
This gives the Dawson function $D_+$ (2.18) again, up to a correction $O(\tau)$
$$I_1=2(1-\tau\d_t){1\over\sqrt{at}}D_+\B({r\over 2\sqrt{at}}\B).\eqno(3.25)$$
So for small thermal relaxation time $\tau$ we recover the term in the solution (2.20) of the usual heat equation.

For large $\tau$ the integral over $[\,k_0,\infty]$ gives the leading contribution. Since we have two complex conjugate roots (3.14) we obtain a real part
$$I_2=\int\limits_{k_0}^\infty=2{\rm Re}\int\limits_{k_0}^\infty d\,k\B(1-i(4a\tau\,k^2-1)^{-1/2}\B)e^{-{t\over 2\tau}(1-
\sqrt{4a\tau\,k^2-1}}\sin\,k r=$$
$$=2e^{-t/2\tau}\int\limits_{\,k_0}^\infty d\,k\,\B[\cos\B({t\over 2\tau}\sqrt{4a\tau\,k^2-1}\B)+{\sin ({t\over 2\tau}\sqrt{4a\tau\,k^2-1})\over\sqrt{4a\tau\,k^2-1}}\B]\sin\,k r.\eqno(3.26)$$
Here the periodic time dependence indicates the appearance of thermal waves [4] which, however, are damped with a time constant $2\tau$. This damping is essential for the approach to steady state The integral (3.26) is investigated in the next section. 

\section{Some special functions}

According to (3.21-23) we must calculate the integrals
$$I^\pm_1=\int\limits_0^{k_0}dk\B(1\pm(1-4a\tau k^2)^{-1/2}\B)e^{\omega_\pm t}\sin kr.\eqno(4.1)$$
With the new integration variable
$$x=2\sqrt{a\tau}k\eqno(4.2)$$
we get the dimensionless form
$$I_1^\pm={1\over 2\sqrt{a\tau}}\int\limits_0^1 dx\B(1\pm{1\over\sqrt{1-x^2}}\B)e^{-{t\over 2\tau}(1\mp\sqrt{1-x^2})}\sin\B({rx\over 2\sqrt{a\tau}}
\B)=$$
$$={1\over 2\sqrt{a\tau}}I_1^\pm(s,u)\eqno(4.3)$$
where
$$s={t\over 2\tau},\quad u={r\over 2\sqrt{a\tau}}\eqno(4.4)$$
and
$$I_1^\pm(s,u)=\int\limits_0^1 dx\B(1\pm{1\over\sqrt{1-x^2}}\B)e^{-s\pm s\sqrt{1-x^2}}\sin ux=$$
$$=\pm e^{-s}(\d_s+1)S_1(\pm s,u).\eqno(4.5)$$
The remaining integral
$$S_1(s,u)=\int\limits_0^1 dx\,e^{s\sqrt{1-x^2}}{\sin ux\over\sqrt{1-x^2}}\eqno(4.6)$$
can be easily calculated by numerical integration, together with its derivatives. But it seems not possible to write it in terms of known special functions.

With the substitution $y=\sqrt{1-x^2}$ we get the form
$$S_1(s,u)=\int\limits_0^1 dy\,{\sin(u\sqrt{1-x^2})\over\sqrt{1-x^2}}e^{sy}\eqno(4.7)$$
and
$${\d S_1\over\d u}=\int\limits_0^1 dy\,s^{sy}\cos(u\sqrt{1-y^2}).\eqno(4.8)$$
On the other hand
$${\d S_1\over\d s}=\int\limits_0^1 dx\,e^{s\sqrt{1-x^2}}\sin ux=\int\limits_0^1 dy\,{y\over\sqrt{1-y^2}}\sin(u\sqrt{1-y^2})e^{sy}=$$
$$=-{1\over u}\int\limits_0^1{d\over dy}(\cos u\sqrt{1-y^2})e^{sy}dy\eqno(4.9)$$
allows partial integration which brings us back to (4.8), so that we get the following differential equation for $S_1$:
$$u{\d S_1\over\d s}=s{\d S_1\over\\d u}+\cos u-e^s.\eqno(4.10)$$ 

For the approach to steady state we need a bound of $S_1(s,u)$ for large $s$. Such a bound is obtained by means of the confluent hypergeometric function [5]
$$\vert S_1(s,u)\vert<\int\limits_0^1{e^{sx}\over\sqrt{1-x}}dx=2 M(1,{3\over 2},s).\eqno(4.11)$$
Using the asymptotic behavior of $M(a,b,s)$ [5] we get for positive $s>0$
$$\vert S_1(s,u)\vert <\sqrt{\pi}e^s s^{-1/2}(1+O(s^{-1})).\eqno(4.12)$$
The exponential factor is cancelled in (4.5) so that we find a slow approach to steady state with $s^{-1/2}$. The other term with negative $s$ behaves
better
$$\vert S_1(-s,u)\vert < 2M(1.{3\over 2},-s)<{1\over s}+O(s^{-2}).\eqno(4.13)$$
This leads to an exponential decrease in (4.5).

To calculate the thermal wave integral (3.26) we introduce the second special function
$$S_2(s,u)=\int\limits_1^\infty{dx\over\sqrt{x^2-1}}\sin(s\sqrt{x^2-1})\sin ux.\eqno(4.14)$$
This is obtained from (3.26) with the substitution (4.2) again. With the new integration variable $y=\sqrt{x^2-1}$ we get a Fourier - sine integral
$$S_2(s,u)=\int\limits_0^\infty{dy\over\sqrt{y^2+1}}\sin(u\sqrt{y^2+1})\sin sy.\eqno(4.15)$$
The same integral with two cosine or one sine and one cosine function can be expressed by Bessel functions [6], but the integral (4.15) cannot. This might indicate that it is a new special function. As it stands the integral is not well suited for numerical integration. We get a better form by using the Euler substitution
$$\sqrt{y^2+1}=yx+1\eqno(4.16)$$
in
$$J_2^\pm=\int\limits_0^\infty{dy\over\sqrt{y^2+1}}\cos (u\sqrt{y^2+1}\pm sy).\eqno(4.17)$$
This follows from (4.15) by simple trigonometric formulas. With the substitution (4.16) we find
$$J_2^\pm(s,u)=2\int\limits_0^1 dx\,{1\over 1-x^2}\cos\B({ux^2\pm 2sx+u\over 1-x^2}\B)\eqno(4.18)$$
which can be easily calculated by numerical integration. For the special case $u=s$ we obtain
$$J_2^\pm(s,s)=2\int\limits_0^1{dx\over 1-x^2}\cos\B(s{1+x\over 1-x}\B)=$$
$$=\int\limits_1^\infty dz\,{\cos sz\over z}=-{\rm Ci}(s),\eqno(4.19)$$
which is the cosine-integral [5]

To obtain a bound for $J_2^\pm$ we use partial integration again:
$$J_2^\pm=\int\limits_0^\infty{dy\over uy\pm s\sqrt{y^2+1}}{d\over dy}\sin(u\sqrt{y^2+1\pm sy})=$$
$$=\mp{\sin u\over s}+\int\limits_0^\infty{\sin(u\sqrt{y^2+1}\pm sy)\over (uy\pm s\sqrt{y^2+1})^2}\B(u\pm{sy\over\sqrt{y^2+1}}\B)\, dy.\eqno(4.20)$$
This decreases as $1/s$ for fixed $u$ or $r$.

\section{Finite spherical geometry}

In the results of the previous sections the zero-mode $k=0$ (3.16) has played an important role. This mode only appears in the infinite system. To be physically relevant we must check whether the infinite system can be considered as a limit of a large finite system. For this purpose let us assume a large spherical boundary of radius $r_1$ which is kept at a constant temperature $T_{01}$, $r_0$ is the radius of the electrode as before. We require the two boundary conditions
$${\d T\over\d r}(t,r_0)=0,\quad T(t,r_1)=T_{01}.\eqno(5.1)$$
The steady state solution satisfying these conditions is now given by
$$T_1(r)=T_{01}-{b\over r_0r_1}+{b\over 2r_1^2}+{b\over r_0r}-{b\over 2r^2}.\eqno(5.2)$$
At the electrode we have the higher temperature
$$T_1(r_0)=T_{01}+{b\over 2}\B({1\over r_0}-{1\over r_1}\B)^2.$$

The remaining time dependent solution $T_2(t,r)$ is again factorized $=T_3(t)T_4(r)$ (2.10) where $T_4$ is the solution of
$$r^2T_4''+2rT_4'=-k^2r^2T_4\eqno(5.3)$$
with the boundary conditions
$${\d T_4\over\d r}(r_0)=0,\quad T_4(r_1)=0.\eqno(5.4)$$
We transform the equation (5.3) into a selfadjoint form with the substitution
$$y(r)=rT_4(r)\eqno(5.5)$$
yielding
$$-y''=k^2 y\eqno(5.6)$$
and the boundary conditions
$$y(r_1)=0,\quad r_0y'(r_0)-y(r_0)=0.\eqno(5.7)$$
This is a simple standard Sturm-Liouville eigenvalue problem [8] in the Hilbert space $L^2([r_0,r_1])$. It has a discrete spectrum of eigenvalues $k_n$ in contrast to the infinite problem in the previous sections. The number $n=0,1,2,\ldots$ gives the number of zeros of the eigenfunctions $y_n$.

The first boundary condition (5.7) is immediately satisfied by
$$y_n(r)=\sin k_n(r_1-r)\eqno(5.8)$$
and the second condition gives the transcendental equation
$$\tan k_n(r_1-r_0)=-k_nr_0.\eqno(5.9)$$
This equation can easily be discussed graphically. It seems as if $k=0$ were the lowest eigenvalue, but a glance to (5.8) shows that this is not the case because $y_n=0$. To get the eigenvalues analytically we put
$$k_n(r_1-r_0)=(2n+1){\pi\over 2}+\delta.\eqno(5.10)$$
Then (5.9) leads to
$$\tan k_n(r_1-r_0)=-\cot\delta=-{1\over\delta}+{\delta\over 3}+\ldots =-(2n+1){\pi\over 2}{r_0\over r_1-r_0}$$
which for large $n$ gives
$$\delta={2(r_1-r_0)\over (2n+1)\pi r_0}$$
so that
$$k_n={(2n+1)\pi\over 2(r_1-r_0)}+{2\over (2n+1)\pi r_0}+O\B({1\over n^2}\B).\eqno(5.11)$$
In the infinite volume limit $r_1\to\infty$ the first term goes to 0, but the second term does not. There remains a finite gap between $k=0$ and the lowest eigenvalue $k_0$. That means the zero-mode of the previous sections is exceptional and not physical. The reason is that the boundary condition (5.7) at $r_0$ is not fulfilled for all $t$ in the infinite problem.

For the selfadjoint eigenvalue problem we have expansion and completeness theorems [8]. The eigenfunctions $y_n$ for different $n$ are orthogonal and complete in $L^2([r_0,r_1])$. To normalize them we compute
$$\int\limits_{r_0}^{r_1}\sin^2k_n(r-r_1)dr={r_1-r_0\over 2}+O\B({1\over n^2}\B)$$
so that
$$\fii_n(r)=\B(\sqrt{{2\over r_1-r_0}}+O({1\over n^2})\B)\sin k_n(r_1-r)\eqno(5.12)$$
is a complete orthonormal system. The general solution of the finite ablation problem for the hyperbolic heat equation is now given by
$$T(t,r)=T_{01}-{b\over r_0r_1}+{b\over 2r_1^2}+{b\over r_0r}-{b\over 2r^2}+$$
$$+{1\over r}\sum_{n=0}^\infty\B[a_n\fii_n(r)e^{\omega_n^+t}+b_n\fii_n(r)e^{\omega_n^-t}\B].\eqno(5.13)$$
Here the two roots (3.14) appear again with $k=k_n$. For the parabolic equation we have only the terms with $\omega_n^+=-ak_n^2$. To satisfy the initial condition
$$T(0,r)=T_{01}\eqno(5.14)$$
we must expand the function
$${b\over 2r}-{b\over r_0}+r\B({b\over r_0r_1}-{b\over 2r_1^2}\B)=\sum_n(a_n+b_n)\fii_n(r)$$
into a Fourier series. To do so we need the $L^2$-scalar products of $\fii_n$ with the functions 1, $r$, $1/r$ which can easily be calculated. The Fourier coefficients $a_n+b_n$ are of the order $1/n$ which gives a slow convergence of the series. For large $t$ the exponential factors in (5.13) give a rapid convergence.

For the hyperbolic equation we again require the second initial condition
$${\d T(t,r)\over\d t}\B\vert_{t=0}=0\eqno(5.15)$$
which yields
$$b_n=-{\omega_n^+\over\omega_n^-}a_n.\eqno(5.16)$$
In the parabolic case this condition is violated. This seems to be in contradiction to experiments. The reason for condition (5.15) is the finite propagation speed of the heat which follows from the characteristics of the hyperbolic equation (3.8) [7] [9]. The characteristics are given by
$${dr\over dt}=\pm{1\over\sqrt{\eps}}=\pm\sqrt{{\kappa\over\ro c\tau}}.\eqno(5.17)$$

\section{Conclusions}

The approach to steady state for large times is determined by the exponential term $\exp(\omega t)$ with frequency $\omega$ closest to 0 in (5.13). Leaving aside unrealistically large $\tau$, this is given by
$$\omega_0^+=-ak_0^2\approx -a\B({\pi\over 2(r_1-r_0)}+{2\over\pi r_0}\B)$$
where $\tau$ has cancelled. Therefore, in contrast to the infinite geometry in sect.2-4, both equations show the same approach to steady state in the finite system.

For small times the solutions of the two equations differ considerably. One reason for this is the different initial condition (5.15). A second interesting difference has been observed by Lopez Molina et al.[4]. The temperature $T$ at fixed radius $r$ shows ``cuspidal-type'' singularities as a function of time $t$. These are discontinuities in the derivative $\d_t T(t,r)$.  The origin of this phenomenon is the change of branch in the characteristic frequencies $\omega_n^\pm$ which is connected with the appearance of thermal waves. Indeed, according to (3.26) we must consider
$$F(t)={\rm Re}\B(e^{\sqrt{1-x^2}t}\B)=\cases{e^{\sqrt{1-x^2}t}&if $x<1$\cr \cos(\sqrt{x^2-1}t)&if $x>1.$\cr}$$
This function is continuous at $t=0$, but $\d_t F(t)$ makes a jump. The theory of characteristics [9] implies that such a discontinuity travels through the medium with the velocity (5.17).


\begin{thebibliography}{} 

\bibitem{} Gonz\'alez-Su\'arez A, Berjano E, Guerra JM, Gerardo-Giorda L, Computational Modeling of Open-Irrigated Electrodes for Radiofrequency Cardiac Ablation Including Blood Motion-Saline Flow Interaction,PloS ONE 11(3): e0150356, doi:10.1371/journal.pone.0150356

\bibitem{} Cattaneo C.R. (1958), Sur une forme de equation de la chaleur eliminant le paradoxe d'une propagation instantanee, Comptes Rendus 247 (4) 431

\bibitem{} Vernotte P. (1958),Les paradoxes de la theorie continue de l'equation de la chaleur, Comptes Rendus 246 (22) 3154

\bibitem{} Lopez Molina JA,Rivera MJ, Trujillo M, Berjano EJ, (2008) Effect of the thermal wave in radiofrequency ablation modeling: an analytic study, Phys. Med. Biol. 53, 1447-1462

\bibitem{} Abramowitz M, Stegun IA, Handbook of mathematical functions, Dover Publications 1965

\bibitem{} Erdelyi A, Oberhettinger MF, Tricomi FG, Tables of integral transforms, vol.1 and 2, McGraw-Hill, New York 

\bibitem{} Vladimirov WS, Equations of mathematical physics; Moscow, Mir Publishers, 1984

\bibitem{} Coddington E.A., Levinson N., Theory of ordinary differential equations, McGraw-Hill Book Company, Inc. 1955

\bibitem{} Courant R, Hilbert D, Methods of mathematical physics,vol.II, Wiley, New York 1989

\end{thebibliography}
\end{document}